\documentclass{epl}
\usepackage{amsfonts}
\usepackage{amsmath}
\usepackage{amssymb}
\usepackage{graphicx}

\newcommand{\beq}{\begin{equation}}
\newcommand{\eeq}{\end{equation}}
\newcommand{\beqn}{\begin{eqnarray}}
\newcommand{\eeqn}{\end{eqnarray}}
\newcommand{\bearr}{\begin{array}}
\newcommand{\enarr}{\end{array}}

\newcommand{\eps}{\varepsilon}
\newcommand{\eg}{{\tt e.g.}}
\newcommand{\ie}{{\it i.e.}}

\newcommand{\Rg}{R_{\rm G}}
\newcommand{\Ri}{R_{\rm I}}
\newcommand{\wtS}{\widetilde{S}_{BG}}
\newcommand{\wtW}{\widetilde{W}}
\newcommand{\wtOm}{\widetilde{\Omega}}

\title{Boltzmann {\it conjecture}, meta-equilibrium entropy, second law, chaos and irreversibility for
many body systems.}
\shorttitle{Boltzmann entropy, Second Law, chaos and irreversibility}
\author{Piero Cipriani\thanks{E-mail: \email{piero.cipriani@istruzione.it}}}
\institute{
\inst{}{\sl MIUR} -- (Italia)\\
\inst{}{\sl ICRA} -- P.le Repubblica, 10-12, 65122 Pescara (Italia)
}

\begin{document}

\maketitle

\begin{abstract}
A heuristic generalization of the {\it Boltzmann-Gibbs microcanonical
entropy} is proposed, able to describe meta-equilibrium features and evolution of 
macroscopic systems.
Despite its {\it simple-minded} derivation, such a function of {\sl collective parameters} 
characterizing the {\sl microscopic state} of N-body systems, yields, at one time, a statistical 
interpretation of dynamic evolution, and dynamic {\it insights} on the basic assumption of 
statistical mechanics.
Its natural (implicit) time dependence, {\it entails} a {\sl Second Law-like} behaviour and allows
moreover, to perform an {\it elementary} test of the {\sl Loschmidt reversibility objection},
pointing out the crucial relevance of Chaos in setting up {\sl effective}
({\it statistico-mechanical and dynamical}) {\sl arrows of time}.
Several concrete (analytical and numerical) applications illustrate its properties.
\end{abstract}

\section{Introduction}
The definition of an {\sl entropy-like} quantity in terms of dynamical
variables for mechanical systems {\it out of equilibrium},
is a fundamental issue of Statistical Mechanics (SM).
In the last decade, most studies focused on Non Equilibrium Stationary 
States (NESS) of driven systems,\cite{GallaCohen}.
In parallel, however, there has been a {\sl revival},\cite{GoldLeb}
of the investigations on the {\sl seemingly simpler} issue,
dealing with isolated systems, left free to evolve from a {\sl non equilibrium} 
initial state; \ie, on the original concern of Boltzmann: the dynamical {\it interpretation} of the tendence 
of macroscopic systems to {\sl spontaneously relax} to an {\sl equilibrium state},
and why and how {\sl natural evolution} is associated with an {\sl increase} of a {\it
macroscopic state variable}, called {\sl Entropy} in the realm of Thermodynamics (TD),
for which a definition in terms of microscopic phase variables is sought
\cite{GoldLebGarr}. In these latter papers in particular, a modification of the 
{\sl (perfect gas) H function}
is suggested, when the contribution of the interaction
potential to the total energy is not negligible.
These works acknowledge the deep (and generally ignored) Jaynes critical analysis, \cite{Jaynes}, 
on the (ab)use of the identification of the {\sl Boltzmann-H} function with
the (negative of the) thermodynamic entropy, and on the (logically equivalent) statement
that the derivation of any flavour of a {\sl H-theorem} should amount to a {\sl proof}
of the {\it Second Law}.

Here, I present a further attempt, inspired again by the Jaynes thought-provoking analysis,
which, far from giving a {\sl mathematically rigorous} derivation, relies upon a 
{\it na\"\i ve extension} of a well known (and implicitly commonly adopted) generalization of the 
Boltzmann-Gibbs (BG) {\sl microcanonical} entropy and does not require any
modification of the {\sl most simple and fundamental mechanical definition
of entropy},\cite{Gross}.
The interconnections with
other recent works {\sl around} this area, either from the viewpoint
of an axiomatic treatment  of  the Second Law, \cite{LiebYngv},
or dealing with NESS and related {\it chaotic hypotheses},
\cite{GallaCohen,Galla} will appear\footnote{
This paper start to summarize studies and {\sl reflections}, 
formed during my Ph.D. studies,\cite{CPTD}.
The {\sl revival} of the investigations on the dynamical
foundations of the Second Law and the wish to acknowledge the conceptual relevance 
of the work of E.T.Jaynes, persuaded me, eventually, to collect those old notes. 
} elsewhere,\cite{CP_PRE}, where conceptual, analytical and numerical aspects
are also discussed in more details.

In the following, I describe the basically simple idea, and how it, leaving formally unchanged
the BG entropy, naturally leads to an implicitly time-dependent quantity.
Then I apply it to three different models, discussing {\bf a)} the analytic predictions;
{\bf b)} the {\sl surprising agreement} between these and the 
numerical outcomes, in {\it equilibrium} and {\it non-equilibirum} cases; 
{\bf c)} the {\it experimental} increase of {\it generalized entropy} along dynamical trajectories
and {\bf d)} the evidence of the {\sl crucial} relevance of {\it Chaos} for the onset of
{\it irreversibility}.
\section{The basic assumption}
The {\sl minimally biased} choice for a reconciliation between Dynamics,
{\sl spontaneous evolution} and statistical description of {\it isolated systems}, has to be 
{\sl microcanonical in spirit}: I will deal with N-body systems (in a $f-$dimensional configuration
space), governed by a Hamiltonian:
\beq
H({\underline{q}},{\underline{p}}) = \sum_{i=1}^{N} \frac{{\bf p}_i^2}{2m_i} + 
\sum_{i=1}^{N-1}\sum_{j=i+1}^{N} {\cal V}_{ij}(\| {\bf q}_i - {\bf q}_j\| ) =
E\equiv {\cal K} + {\cal U}\ .
\label{eq:Hamiltonian}
\eeq
The SM over the $(2fN-1)$-dimensional constant energy surface, $\Sigma_E$,
establishes the logical connection between Dynamics and TD, exploited
through the BG entropy,
$S_{BG}(E,N,V) = k_B \ln{W(E,N,V)}$, 
where $W$ is the {\it measure} of $\Sigma_E$ ($k_B$ being the Boltzmann's constant),
and depends on few macroscopic parameters ($E,N$ and, \eg, the spatial volume $V$).

However, it is well known that, in the case of isolated systems, dynamical trajectories actually explore only a (zero-measure)
subset of $\Sigma_E$, liyng down rather on the $D\doteq (2fN-1-M)$-dimensional manifold,
$\Omega$, defined by $E$ and the other $M$ {\sl integrals of motion} of the system.
Indeed, if a consistent SM is sought, $S_{BG}$ has to be suitably modified, whenever the dynamics admits additional 
{\sl conserved or costrained quantities}.
This generalization is simple and is {\sl tacitly assumed} in most textbooks
and discussions, mainly because, generally, it is $M\ll N$ and the results do not depend on the choice
(see, however \cite{Padma} for a curious counterexample).
The above definitions lead to the microcanonical SM and work egregiously well for equilibrium TD.
However, it has been argued (see, \eg,\cite{EE}) that the above expressions, if suitably interpreted, 
could give a measure of entropy even in nonequilibrium states. Although such a terminology will horrify
anyone eager for rigorous definitions, Jaynes instead was a convinced supporter of this interpretation, 
though referring to the {\it Gibbs entropy}, $S_G = -k_B \int_\Omega \rho\ln{\rho} d\Gamma$, \cite{Jaynes}.

In the same spirit, the {\it basic extension} here proposed consists in assuming that the region $\wtOm $ 
whose measure, $\wtW\doteq \|\wtOm\|$, enters in the definition
of the generalized entropy, depends not only on the (conserved)
values of energy, $H({\underline{q}},{\underline{p}})\equiv E$ and on the additional integrals of motion, 
${\bf K}({\underline{q}},{\underline{p}})\equiv{\bf I}$, 
where ${\bf I}\doteq\{I_1,\ldots,I_M\}$, fixed {\it \`a priori} by the initial conditions and 
thereafter  {\sl rigorously} conserved, but can have a functional dependence also on other
collective variables, $\{{\cal P}_i\}$, whose evolution is instead determined
(apart from statistical fluctuations), implicitly and self-consistently, 
by the dynamics. 
As the $\{{\cal P}_i\}$ evolve, according to Hamilton's equations,  the measure of the {\it effectively
available phase-space volume} changes, so that the {\it generalized entropy function}, $\wtS$, of the system 
is allowed to evolve as well: were $\Omega=\Omega(E,N,{\bf I})$ alone, then, clearly, no entropy evolution could 
be {\it explained}.

{\it The basic consistency requirement amounts to verify whether the {\sl quasi deterministic} evolution of the $\{{\cal P}_i\}$
implies {\sl almost always} an increase of $\wtS$, {\bf and}, viceversa, whether the maximization of the
latter with respect to the macroscopic parameters, gives predictions about their behaviour consistent with the dynamics.}
In the affirmative case, a meaningful (yet heuristic) reconciliation between Dynamics and TD is obtained.
It has to be emphasized how, in this scheme, {\sl physical intuition} plays a major role to single out the best suited 
macroscopic parameters able to describe the dynamical and statistical evolution of the system. All selection criteria 
should consider that TD and SM, deal with {\sl macroscopic} systems; consequently, a possibly singular behaviour of single 
microscopic {\sl constituents}, which {\it does not alter} the values of collective parameters, cannot modify the 
pertinent level of description.
To show the effectiveness of the approach, I will confine to few (yet {\sl physically relevant}) 
cases, where the above selection is more or less straightforward and the physical meaning 
of the parameters transparent\footnote{In particular, dealing with {\sl self-confining}
interparticle potentials allows to neglect the effects of a {\sl vessel} and to exploit the fact that, once 
$E$ and $N$ are fixed, the spatial volume $V$ is implicitly costrained within well defined limits, 
at least on timescales of interest, see\cite{CP_PRE}.
}.
Therefore, introducing the specific energy, $\eps\doteq E/N$, and assuming that the $\{{\cal P}_i\}$ vary within suitable 
intervals $\Pi_i(\eps,N;{\bf I})$, the generalization suggested reads:
\beq
\wtS (\eps,N,{\bf I};\{{\cal P}_i\}) \doteq
k_B \ln{\widetilde W} = k_B \ln\int_{{\cal P}_i\in\,\Pi_i}\! \delta[E-H({\underline q},{\underline p})]\, \delta^{(M)}[{\bf I}-{\bf K}
({\underline q},{\underline p})] 
{d{\underline q}d{\underline p}}\ .\label{eq:W}
\eeq
The (implicit) time dependence is hidden in the dynamically determined evolution
of the $\{{\cal P}_i\}$.
Such {\it macroscopic variables} can be often singled out analyzing the scaling properties of the model. A common example,
exploited below, is the {\it virial ratio}, $Q\doteq A_{\cal V}\langle {\cal K}\rangle/\langle {\cal U}\rangle$,
between the (time averages of) kinetic and potential energies. A purely dynamical theorem (see, \eg\cite{Clausius,GoldsteinH}) 
assures that it attains, for confined motions, a well defined value, 
dependent, in the general case, only on the form of potential ${\cal V}$ and, possibly, on $\eps$. The phase-space volume can often
be expressed as a function of the related {\it instantaneous virial ratio}, ${\cal Q}(t)\doteq A_{\cal V} {\cal K}(t)/{\cal U}(t)$.
\section{Applications}
As a first example let us consider the simplest N-body hamiltonian system
(except, {\it perhaps}, the perfect gas): a chain $(f=1)$ of N harmonic oscillators with nearest neighbour interactions. 
Although this system is integrable, for irrational frequency ratios, {\it kinematic phase mixing} takes place and 
motions of single oscillators decorrelate. 
It is obvious that no irreversible evolution can occur; however, while some variables
approach  a definite value only in {\sl time averaged} sense, others assume, 
{\it even locally}, values practically coincident  with 
asymptotic ones. For example, in this case, $Q_{ho}\doteq\langle {\cal K}\rangle/\langle {\cal U}\rangle \equiv 1$
is energy independent and ${\cal Q}(t)$ oscillates around this value, within statistical ${\cal O}(N^{-1/2})$ fluctuations, 
unless collective oscillations occur, which are, however, quickly damped out by the (kinematic) mixing.
As a first {\sl exercise},\cite{CPTD}, I show how the dynamic {\sl evidence}, ${\cal Q}\cong 1$, receives within the present
approach a statistical {\it explanation} (or, at least, {\it interpretation}). 
Indeed, assuming for simplicity $m_i\equiv m$
 and introducing suitable scales, $(R,V)$, for positions and momenta coordinates,
${\bf q}_i = R {\bf x}_i$, 
and 
${\bf p}_i = mV {\bf y}_i$,
such that $2\eps = mV^2 + {m}\overline{\omega^2}R^2$,
with $\overline{\omega^2}$ a {\sl weighted average frequency} and exploiting the definition of 
${\cal Q}(t)$, it is possible to write,\cite{CP_PRE},
\beq
V^2 = \frac{2{\cal Q}\,\eps }{m(1+{\cal Q})}\ ;\ R^2 = \frac{2\eps}{m\overline{\omega^2}(1+{\cal Q})}\ ,
\eeq
from which, substituting into eqs.(\ref{eq:W}), we obtain
\beq
\wtS = \frac{fN}{2}k_B \ln\left[ \frac{4{\cal Q}\,\eps^2}{\overline{\omega^2}(1+{\cal Q})^2}\, 
{\cal I}({\underline x},{\underline y})\right] \ ; \label{eq:Sho}
\eeq
with ${\cal I}({\underline x},{\underline y})$ a dimensionless integral, depending 
only on the {\it microscopic} state, and evolving (much) more slowly than, ${\cal Q}$.
Despite its simplicity, eq.(\ref{eq:Sho}) has two important consequences:
{\sf a)} the maximization of $\wtS$ with respect to ${\cal Q}$, yields the {\sl
virial equilibrium} condition: $\partial \wtS /\partial {\cal Q} = 0\ 
\Longleftrightarrow {\cal Q}=1$; {\sf b)}  the microcanonical temperature, 
$T\doteq (\partial \widetilde{S}_{BG}/\partial E)^{-1}$, leads to the
well known formula for the heat capacity of a (1-D) harmonic crystal:
$E = Nk_B T$, coherent, incidentally, with the {\it orthodicity} requirement,\cite{GallaBook}.

Before to go beyond what could have been considered just a funny {\sl educational
curiosity}\footnote{
And I've been so persuaded fifteen years ago, even if it was not 
the only example I had found at that time.
},
I want to point out that, in the general case, the evolution of the {\sl collective} parameters is not the
only source of entropy increase; however, a separation of scales (of  size and
time) exists, such that the {\sl gross and faster} evolution is driven by these
quantities, \cite{CPMP,CP_PRE}.

Let us consider now a system (called here RMLB-model) which has been used to model phase transitions in small 
clusters of atoms,\cite{RMLB}, whose Hamiltonian is
\beq
H({\underline{q}},{\underline{p}}) = \sum_{i=1}^{N} \frac{{\bf p}_i^2}{2m_i} + \frac{1}{2}
K_0 N r^2\left[ 1-br^2+cr^4\right]\ ,\ \ \ {\rm with}\ \ \ r^2\doteq N^{-1}\sum_{i=1}^N {\bf q}_i^2\ .
\label{eq:RMLBHam}
\eeq
The TD (and, surprisingly, the dynamics too)
of this model presents (even in the case $K_0=1$, $c=1$ and $m_i\equiv 1$) several interesting aspects,
that have been,\cite{RMLB}, or will be discussed elsewhere in details,\cite{CP_PRE}.
The approach proposed, allows to recover its TD
peculiarites and non trivial phenomenology (as, \eg, multiple local entropy maxima
and the occurrence of {\it negative heat capacities}) and
to {\it interpret} the connections between dynamical and TD computations, even in a system, again,
without any {\sl dynamic stochasticity}.

\begin{figure}
\twofigures[scale=0.6]{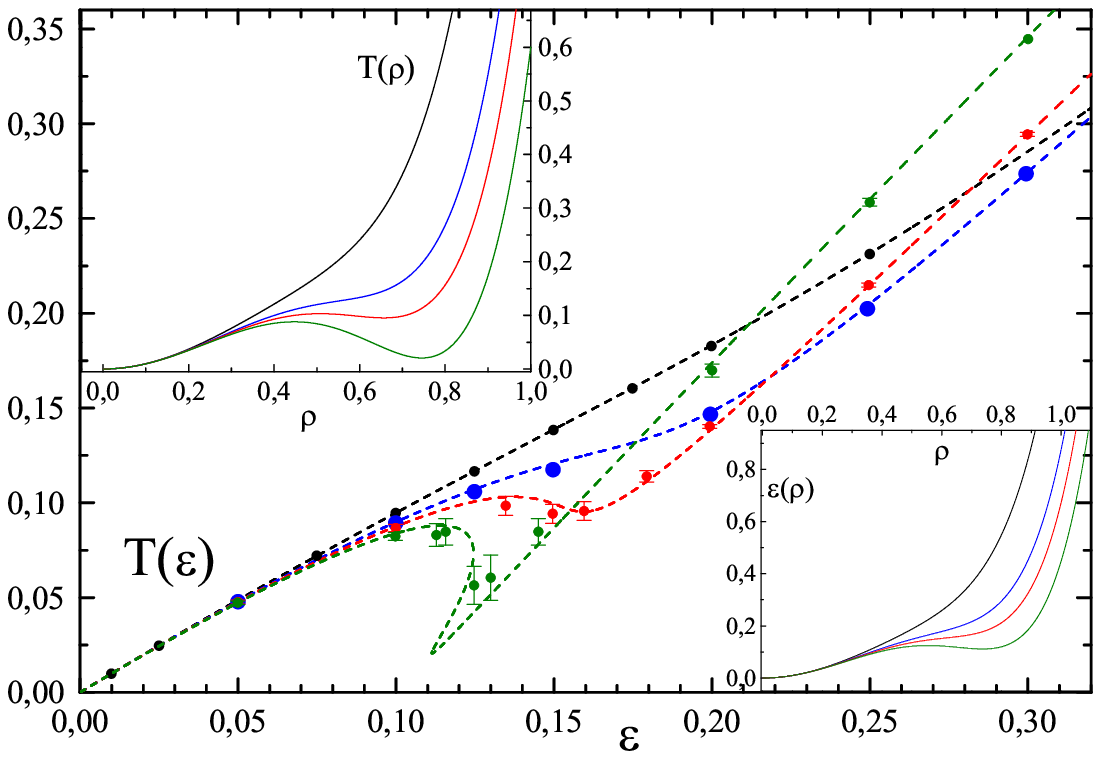}{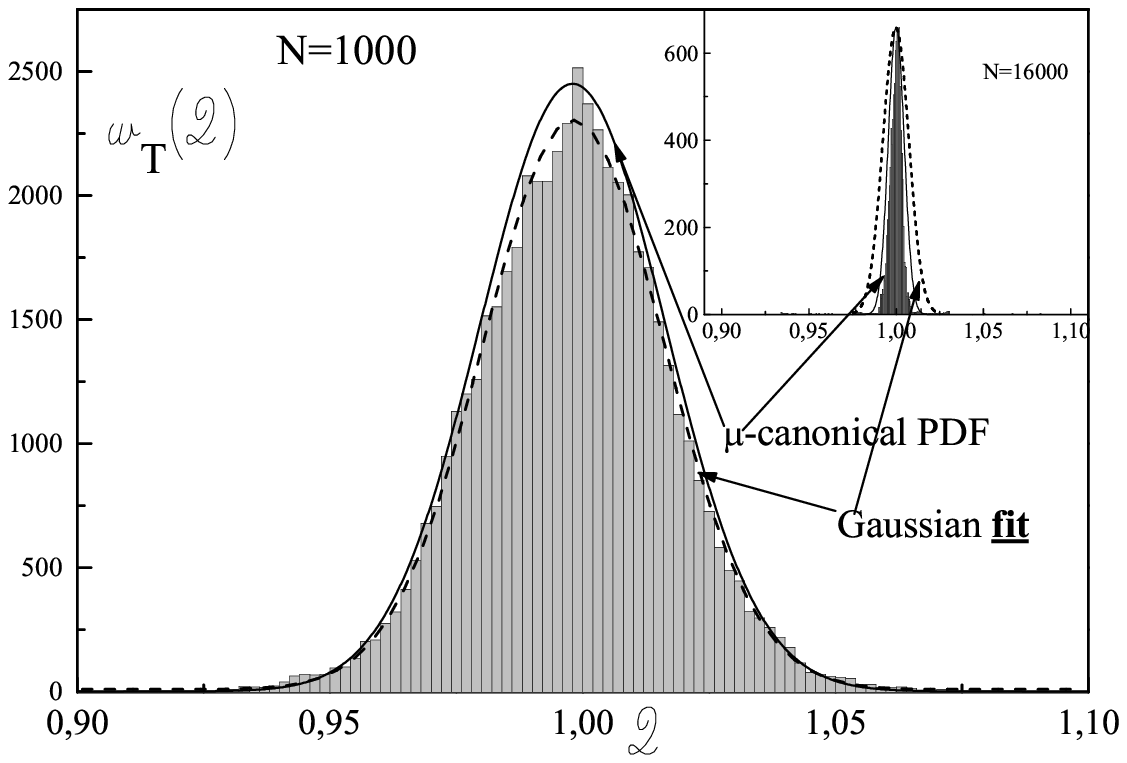}
\caption{Comparison of the SM (lines) and dynamical (symbols) caloric curves of the RMLB model. 
Black, blue, red and green curves refer, respectively, to: $b=1.0\ ;\ \sqrt{2}\ ;\ 1.55\ {\rm and}\ 1.7$.
The insets show the behaviour of $T(\rho)$ and $\eps(\rho)$.}
\label{fig:RMLB}
\caption{Comparison of the theoretical, microcanonical distribution of the ${\cal Q}$-values,
predicted from eq.(\ref{eq:SNBG}), with the {\sl experimental}, {\it relaxed} distribution, computed along a
numerically integrated hamiltonian trajectory, for a SGS with $N=1000$. The dashed curve represents
a gaussian {\sf fit}. In the inset the same comparison is made, for $N=16000$ but for a much shorter time.}
\label{fig:VRdistribution}
\end{figure}

Assuming, as before, $f=1$, exploiting, in this case, the {\sl virial theorem},\cite{CP_PRE},
and introducing the specific entropy, $\sigma\doteq\wtS/Nk_B$, we get a {\it parametric 
entropy-energy relation} through the {\sl order parameter} $\rho\doteq\langle r^2\rangle^{1/2}$:
\beq
\eps =(\rho^2 - \frac{3b}{2}\rho^4 + 2\rho^6)\ ;\ \ 
\sigma\cong\frac{1}{2}\ln\left[2\rho^2 \left(\rho^2 - 2b\rho^4 + 3\rho^6\right)\right]\, ,
\label{eq:eps_sigma_dirho}
\eeq
As before, the microcanonical Temperature, which reads here: 
$k_B{\displaystyle T(\eps ) = \left.\frac{(d\eps/d\rho)}{(d\sigma/d\rho)}\right|_{\rho=\rho_{\eps }}}$
(with $\rho_x$ such that $\eps (\rho_x)= x)$),
yields the caloric curves, and confirms also for this model, that, within this approach,
the {\it orthodic condition} holds,\cite{CP_PRE}.
To compare concretely the predictions of the proposed SM with the outcomes of Dynamics, I've performed several  
Molecular Dynamics (MD) simulations of Hamiltonian (\ref{eq:RMLBHam}) 
for different values of $b$ and $\eps$ and computed the corresponding {\sl dynamical temperatures}, through well 
known formulas,\cite{Pearson}. These are compared, in fig.\ref{fig:RMLB}, with the {\sl theoretic} curves derived above. 
We see that the agreement is absolute almost everywhere, except in the regions of {\it negative heat capacity}. 
As anticipated above, the interesting features of this 
model, indeed, originate from the existence of a threshold value, $b>b_*$, above which coexistence regions appear. 
There, as shown in the insets of 
fig.\ref{fig:RMLB}, both $\eps(\rho)$ and $\sigma(\rho)$ are not anymore invertible functions, so that $\rho_\eps$
is multivalued as is $T(\eps)$. This is not at all a drawback: on the contrary, 
the big fluctuations there occurring, help to understand the nature of the {\sl coexistence} in this 
system: in a (microcanonical) MD simulation $\eps$ is fixed, though this doesn't fix obviously
$r$. In a {\sl pure phase}, $b<b_*$, however, it oscillates in a {\it single} range,
so that $\rho$ attains a well defined value; if $b>b_*$, instead, $r$
can be {\sl trapped}, intermittently, within different {\sl bands}, compatible with the given $\eps$. 
Correspondingly, $\rho$ and, consequently, $\sigma$ and $T$, show much larger fluctuations.

The dynamics of both models analyzed so far is not {\it enough unpredictable}, and this makes them
not exactly the kind of system one expects to be suited for a SM approach. Despite this, the results shown
allow to argue that {\sl deterministic chaos} is not always essential for a (at least partially)
successful SM description of (meta-)equilibrium states. 

To investigate deeply the distinguishing features of TD, \eg the Second Law and the origin of an {\sl effectively irreversible evolution},
I will focus now on the (3D) self-gravitating N-body system (SGS), whose dynamics is surely unpredictable, has been the original 
{\it stimulus} which brought about,\cite{CPTD}, the first {\it reflections} and for which preliminary accounts have already appeared,\cite{CPMP}.
For simplicity, I assume to have a {\it bound} ($\eps\equiv E/N < 0$) system of N equal mass ($m_i\equiv m$) particles, interacting via 
a softened\footnote{
As discussed elsewhere, \cite{CP_PRE}, the softening is adopted here not (only) to avoid numerical singularities, but mainly 
to be fully coherent with the basic assumptions of the approach proposed. Therefore, in what follows
 $d_{ij}\doteq \sqrt{(r_i - r_j)^2 + \eta^2}$, where 
$\eta\doteq\eta_* \overline{d}$. That is, the two-body interaction is {\sl softened}
when the particles get closer than a (small) fraction of the mean interparticle separation,
$\overline{d}\equiv GN^{2/3}m^2/|\eps|$.
In all the numerical simulations here discussed, every particle has unit mass, the initial conditions 
are generated at {\it virial equilibrium}, $-2{\cal K} = {\cal U}$, with specific energy fixed $\eps=-10$, 
and particles distributed initially with uniform spatial density in a spherical box of radius $\Rg$, 
with a velocity distribution approximately maxwellian. Time is measured
in units of {\it dynamical (or crossing) time}\cite{BT,CPTD,CPMP}, $t_D\doteq (G\varrho)^{-1/2}\cong 
N/|\eps |^{3/2}$. The numerical integrations are performed in double precision, using a high-order symplectic algorithm,\cite{McLA}, 
and the time-step is chosen small enough to keep the (maximum) relative energy error below $10^{-4}$. 
The position and the velocity of the center of mass of the system are rigorously conserved within the 
round-off errors, \ie, within $10^{-14}$.
The consequences (or lack thereof) of modifying some of the choices are discussed elsewhere,\cite{CP_PRE}.
}
Newtonian potential, $V_{ij}(\| {\bf q}_i - {\bf q}_j\| ) = -G m_i m_j / d_{ij}$.
Gravitational interaction is not rigorously {\it confining}; it can be argued, however,\cite{BT,CPTD,CPMP} that, up 
to any physically meaningful time, the number of {\it escaping particles} is negligible. Thus, as before, we can 
introduce suitable scale factors for coordinates and momenta, defining
the {\it gravitational} and {\it inertial} radii, 
$\Rg\doteq (\sum m_i)^2/(\sum_{i\neq j}m_i m_j/d_{ij})\approx GNm^2 /|\eps |$
and $\Ri\doteq\left[\sum m_i r_i^2 / \sum m_i\right]^{1/2}$, respectively, and the ratio
among them, $\alpha\doteq\Ri /\Rg$. 
After some elementary algebra,\cite{CP_PRE}, we can express again the phase-space volume
as a function of {\sl standard} parameters $(\eps,N)$, {\sl collective} phase functions, 
${\cal Q}\doteq -2{\cal K}/{\cal U}$ and $\alpha$, 
and a dimensionless integral, 
${\cal J}$, accounting for the correlations amongst particles, to obtain:
\beq
\widetilde{S}_{BG} \doteq Nk_B\sigma\equiv \frac{3Nk_B}{2} \ln\left[-\frac{G^2N^2m^5}{2\eps}\,\alpha^2 {\cal Q}\left(2-{\cal Q}\right)\,
{\cal J}({\underline x},{\underline y})\right] \ . \label{eq:SNBG} 
\eeq
The different kinds of parameters in eq.(\ref{eq:SNBG}) play distinct roles: as far as we consider the system isolated,
$\eps$ and $N$ (and clearly $m$) are rigorously constant, whereas ${\cal Q}$, $\alpha$ and ${\cal J}$ evolve, though
on very well separated (and increasing) timescales. A thoroughful analysis of their hierarchy is presented elsewhere, \cite{CP_PRE,CPMP}.
The implications of eq.(\ref{eq:SNBG}) follow immediately: {\bf a)} the definition of the {\it microcanonical temperature}
leads to the {\it caloric curve} for SGS,
\beq
T^{-1}\doteq (\partial\widetilde{S}_{BG} /\partial E)\equiv k_B (\partial\sigma/\partial\eps)\ \Longrightarrow \eps = -\frac{3}{2}k_B T\ ,
\eeq
yielding a direct proof of the well known (for SGS) {\sl negative heat capacity}. {\bf b)} As in previous cases, $\widetilde{S}_{BG}$
in eq.(\ref{eq:SNBG}) is maximized for ${\cal Q}=1$, and this confirms again the statistical intepretation of virial theorem.
{\bf c)} What's more, for this system, the present approach allows to go well
beyond the {\it mere} prediction of the correct average value: eq.(\ref{eq:SNBG})
provides also a {\sl distribution law} for the values of ${\cal Q}$ in the microcanonical ensemble. 
This is a rather convincing result if we ponder over the surely chaotic nature of SGS,\cite{CPTD,CPMT_pss,CPMP}; 
implying that their statistical properties are likely to be very strong.
It is therefore reasonable to expect that (microcanonical) ensemble probability 
distributions should be {\it accurately sampled} during a long enough hamiltonian trajectory.
The agreement between Dynamics and (modified) microcanonical SM is clearly
evident from Figure \ref{fig:VRdistribution}: the histogram reports the effective 
counting of the actual values assumed by ${\cal Q}(t)$ along a numerically integrated trajectory 
($N=1000$ and $\eta_*=0.01)$) while the solid line represents (up to a normalization factor) the {\sf pdf} for 
the ${\cal Q}$-values, as predicted by eq.(\ref{eq:SNBG}).
It can be argued that both distributions are very close to the gaussian curve (dashed line), which seems to reproduce equally well the 
numerical data. However, there is a {\it physically important} distinction: the gaussian
results from an {\it optimized} empirical {\it fit}, whereas the $\mu$-canonical {\sf pdf} has no fitting parameters. 
The difference is strinkigly evident in the inset, where an analogous comparison is made for 
a simulation with $N=16000$, but on a much shorter time, just comparable with the {\it virialization time}. 
There, the gaussian fit is apparently unsatisfactory, as the {\it weight} of the (relatively) large 
oscillations of ${\cal Q}$ during the {\sl virialization process}, imposes to the {\sl fit} a very large width. 
On the other hand, it should be stressed that, on increasing $N$ ({\sf and} awaiting for a long enough 
integration time), the $\mu$-canonical {\sf pdf} and the gaussian fit will overlap, as the
former, in the $N\rightarrow\infty$, approach a gaussian with width $\sigma^2\propto N^{-1}$ and
the amplitude of the fluctuations will vanish, so that the gaussian fit shrinks,\cite{CP_PRE}.
Thus, the proposed framework, allows to reproduce the statistical distribution of collective 
parameters along hamiltonian trajectories even for finite time, when equilibrium distributions fail, 
recovering, however, in the $N\rightarrow\infty $ and $t/t_D\rightarrow\infty $ limits, the {\sl normal}
equilibrium results.
\begin{figure}[t]
\twofigures[scale=0.6]{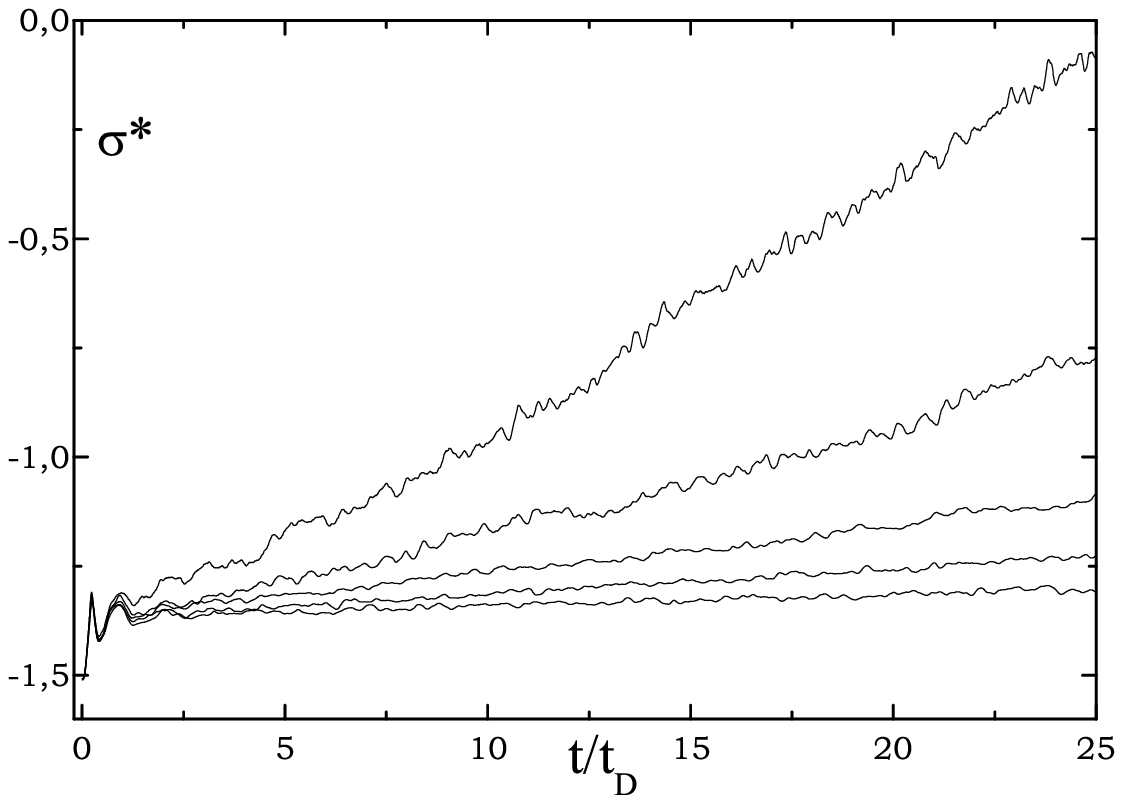}{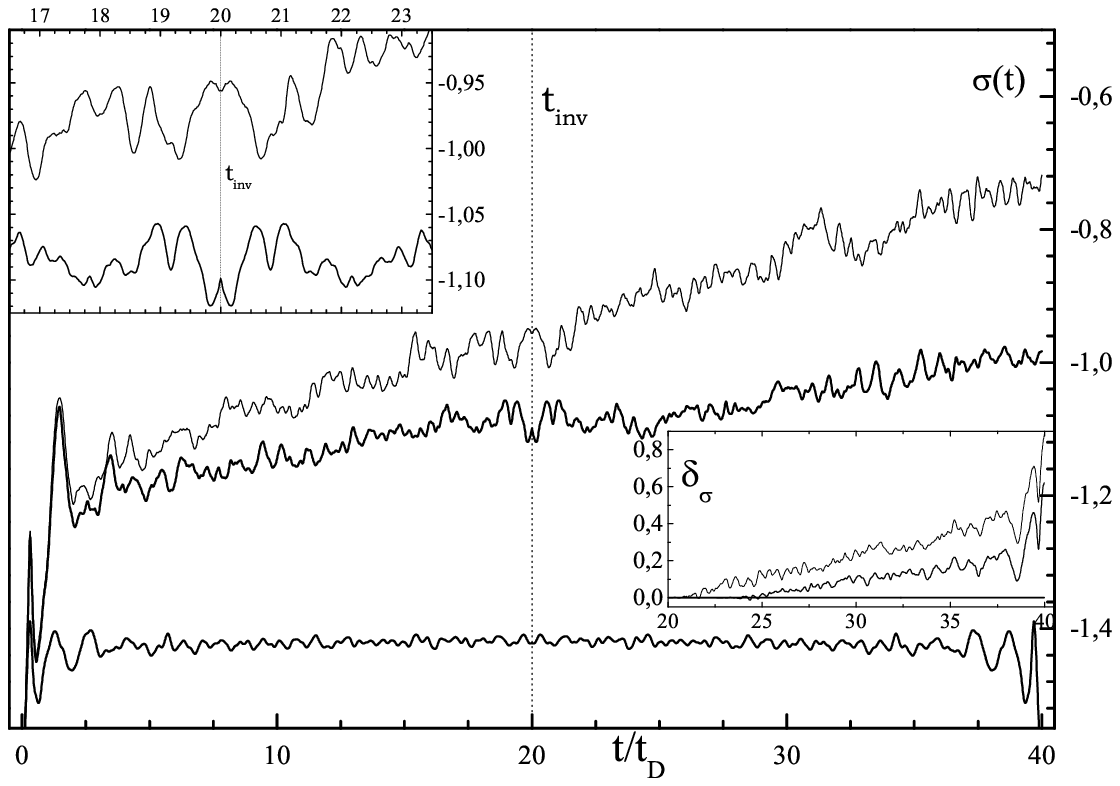}
\caption{Increase with time (measured in units of the {\it dynamical time}, $t_D$)
of $\sigma^*$, along hamiltonian trajectories of
several SGS. From top to bottom it is $N=$ 1000, 2000, 4000, 8000, 16000. Except for the largest $N$,
the plots represent the average of different {\it realizations}. Always it is $\eta_*=0.02$.}
\label{fig:NBsigmadit}
\caption{{\it Reversibility objection and Chaos.} Evolution of $\sigma^*$ along single
trajectories of $N=4000$ SGS's with $\eta_* = 0.02,\,0.2,\,2$, from top to bottom.
At $t=t_{inv} = 20t_D$ the velocities are reversed. The upper left inset is an enlargement around $t_{inv}$
for the smallest $\eta_*$; the other inset contains the plots of $\delta^* (t)=\sigma^*(t)-\sigma^*(2t_{inv}-t)$.}
\label{fig:NBreversibility}
\end{figure}
So far we have thus seen that a {\it maximum entropy}-like criterion applied to
$\widetilde{S}_{BG}$ is not only consistent with dynamical equilibrium 
conditions but predicts also correctly (meta-)equilibrium statistical and TD properties 
of many body systems. However noteworthy may be that, one could wonder whether $\widetilde{S}_{BG}$ 
satisfies also deeper properties expected to hold for an {\it entropy} function. 
If we were able {\sl to show} (if not {\sl to prove}!) that $\widetilde{S}_{BG}$ {\it increases} almost monotonously 
with time, then we could be really confident that the {\it na\"\i ve}
estimate of the phase space volume $\widetilde{\Omega}(E,N,{\bf I};\{ {\cal P}_i\})$ contains a profound meaning,
connected with the {\it Second Law} and the issue of irreversibility.
That this is indeed the case is, I think, unambiguously shown in figure \ref{fig:NBsigmadit}, 
where the (numerical) time evolution of the quantity\footnote{
Which is, essentially, the only evolving contribution to the {\it specific entropy} $\widetilde{S}_{BG}/Nk_B$, 
as the other time-dependent quantity, the integral ${\cal J}$, evolves on much longer timescales.
Notice that time is measured in units of the so-called {\it dynamical time}, $t_D\propto N$.
}
$\sigma^*\doteq \ln\left[\alpha^2 {\cal Q} \left(2-{\cal Q}\right)\right]$,
for different values of $N$. 
After a relatively fast (with a $N$-independent rate) initial increase,
$\sigma^*$ steadily grows, although with a slower rate the larger is $N$,\cite{CP_PRE}, and with fluctuations
which also are smaller the greater is $N$\footnote{
The details and the implications in astrophysical context of these results
will be discussed elsewhere, along with the clear-cut {\it intensive nature} of $\sigma^*$ 
emerging from the figure.
}: $\widetilde{S}_{BG}$, computed along (microcanonical) dynamical trajectories behaves like
the TD entropy and, moreover, the {\it frequency of local (in time) violations of the Second Law} vanishes
for large $N$.

This is {\it much of}, though not all, the content of {\sl the Second Law}: this law establishes
indeed what is (popularly) known as an {\it arrow of Time}; that is, it enables to distinguish the {\sl past}
from the {\sl future} evolution. This contrasts with Hamiltonian dynamics, {\sl invariant under time reversal}. 
I will not discuss here\cite{CP_PRE} the many attempts to 
settle the issue, but simply recall two amongst the many arguments invoked to justify the onset
of macroscopic irreversibility in the dynamics of mechanical systems:
the occurrence of {\it non typical initial conditions} and
the {\it presence of a chaotic dynamics}.
Indeed, discussing the first two models, we have seen that even systems which are definitely 
not chaotic, possess (some) collective properties which can be described {\it statistically}. This
amounts to say, that TD concepts can be useful to describe at least some {\it equilibrium properties} of 
even {\it dynamically regular} macroscopic systems, 
provided that no reference is made to non equilibrium behaviour, associated to the {\it Second Law}.

With regard to the issue of irreversibility, were the entropy increase due {\it solely} to the peculiarity of the 
initial state, then the {\it Loschmidt reversibility objection} would not lead to any paradox, and entropy could
possibly decrease {\it going back in time}. I claim, however, that this is not the case, and precisely that
Chaos plays an essential role in the true onset of irreversibility. In this perspective, therefore,
the relationship between {\it Second Law} and irreversibility loses a tiny part of its complete and
absolute logical equivalence: the first can originate from {\it a non-generic initial state} of the
system, the impossibility to retrace exactly the evolutive path backward in time arises from the presence
of a chaotic dynamics. In this evenience, consequently, {\it entropy  functions} exist that  increases
in both directions of time, irrespectively of the choice of initial conditions.

Figure \ref{fig:NBreversibility} unequivocally {\sl supports} the above assertions, giving them a
clear {\it empirical evidence}.
The curves there shown represent, as in fig.\ref{fig:NBsigmadit}, the time evolution of $\sigma^*$ for three $N=4000$ 
particles systems, all {\it starting from the same initial conditions}, but with varying softening parameter, $\eta$. 
At variance with the previous figure, however, the validity of {\it Loschmidt reversibility objection}\footnote{
It is wise to recall that Loschmidt's objections (as well as Zermelo's one) were against the Boltzmann's
{\it H-theorem}, so related to the interpretation of the Boltzmann's $H$-function, based on the
single particle distribution function. 
Nevertheless, it is clear that Loschmidt's criticisms could apply equally to the
present context (whereas for the Zermelo's objection it is not exactly equivalent,\cite{CP_PRE}).
} is checked, reversing, at time $t=t_{inv}$, every particle's velocity. 
The results are clear: {\bf A)} for a strongly, or even moderately, chaotic system, the {\it entropy evolution} is 
reversed only for very short time intervals after the inversion; subsequently the (dynamical) 
instability drives again the {\sl collective variable} $\sigma^*$ to increase, obeying to a {\it Second Law} even for
the backward time evolution (see upper inset).
{\bf B)} Viceversa, if the dynamics is {\it virtually regular}, as in the case of a very large $\eta_*$, although
an initial entropy increase occurs as well, due to the {\it non complete genericity} of the initial state,
after the inversion, the {\it macroscopic} quantity $\sigma^*$, traces back along its path, taking
{\it exactly the same} values assumed prior to inversion, and accompanies the return of the 
system to the initial (macroscopic and microscopic) state (lower inset).

I'm obviously aware that the above results deserve a deeper discussion\cite{CP_PRE} and that the interpretation of the
numerical effects mimicking the {\it unavoidable external disturbances} should be carefully investigated; nevertheless, 
I find it extremely instructive, as it shows a clear link between microscopic instability (let it be driven by
{\it numerical or ambiental noise}) and irreversible behaviour of macroscopic quantities, whether it is agreed upon
to call them {\it entropies} or not.

\section{Conclusion}
A simple {\it na\"\i ve} estimate of phase space volume, compatible with the values of macroscopic
parameters of several hamiltonian N-body systems has been used to introduce a simple generalization of the BG entropy that
obeys and satisfies the corresponding {\it Second Law}.
The results here presented are unambiguous and do not require anything more than an undergraduate
mathematics. The same conceptual approach can (and will\cite{CP_PRE}) be generalized to a wider class of systems,
just using a more refined formal treatment. Admittedly, the {\it extreme simplicity} and the heuristic nature
of the approach call perhaps for a more {\sl rigorous} formalization. Nevertheless, its relevance 
resides just in its {\it elementary} derivation and discloses itself through the ability to 
describe {\it adaptively} the TD properties of a system in terms of suitable 
macroscopic parameters which evolve on suitable timescales.
Another conceptually appealing feature of the approach here proposed is that
the hypotheses needed for different levels of statistical descriptions
emerged naturally, along with the development of the reasoning.
In particular, we have seen how the conditions, like {\sl literal ergodicity} 
or {\sl mixing}, usually required, are not always essential
to justify an effective (though {\it partial}) TD description.

\end{document}